# Virtual teaching assistant for undergraduate students using natural language processing & deep learning ⊘

Sadman Jashim Sakib ✉; Baktiar Kabir Joy; Zahin Rydha; Md. Nuruzzaman; Annajiat Alim Rasel

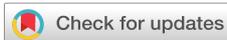

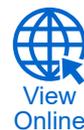 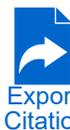 CrossMark

View Online

Export Citation

25 February 2024 05:24:11

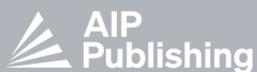

# Virtual Teaching Assistant for Undergraduate Students Using Natural Language Processing & Deep Learning

Sadman Jashim Sakib[1, a)], Baktiar Kabir Joy[1)], Zahin Rydha[1)], Md. Nuruzzaman[1)] and Annajiat Alim Rasel[1)]

[1] *Department of Computer Science and Engineering, Brac University, Dhaka, Bangladesh*

a) Corresponding author: sadman.jashim.sakib@g.bracu.ac.bd

**Abstract.** Online education's popularity has been continuously increasing over the past few years. Many universities were forced to switch to online education as a result of COVID-19. In many cases, even after more than two years of online instruction, colleges were unable to resume their traditional classroom programs. A growing number of institutions are considering blended learning with some parts in-person and the rest of the learning taking place online. Nevertheless, many online education systems are inefficient, and this results in a poor rate of student retention. In this paper, we are offering a primary dataset, the initial implementation of a virtual teaching assistant named VTA-bot, and its system architecture. Our primary implementation of the suggested system consists of a chatbot that can be queried about the content and topics of the fundamental python programming language course. Students in their first year of university will be benefited from this strategy, which aims to increase student participation and involvement in online education.

## INTRODUCTION

Initially, we had several choices for selecting a training topic for our system. Since we have a background in computer science, we have seen that students struggle to comprehend computer programming owing to a lack of resources. To improve the teaching of any subject, available innovations must be utilized. One of the most important skills for the Fourth Industrial Revolution is programming [1]. The majority of students avoid programming because they perceive it to be difficult. Additional learning and teaching techniques may be employed to aid in the learning process. Undergraduates have less concern about development environments, and the majority believes that hands-on learning is more valuable than studying theory based subject matter [2]. Our research aims to solve the issue of a significant number of new learners who become stuck while learning the fundamentals of the python programming language and may not obtain appropriate assistance through the online-based education systems of many institutions.

## RESEARCH AIM

Our study aims to assist a significant number of students who struggle with a variety of issues and oftentimes become disoriented when adapting to programming-related concepts. Moreover, at certain universities, there may not be enough support to assist students with the online-based education system. Thus, the purpose of this research is to introduce a virtual teaching assistant (VTA) that will facilitate student learning and decrease the possibility of failure. Again, due to the shortage of adequate data in this field, we aim to create a primary dataset for our system. So now the chatbot can automatically respond to a variety of inquiries pertaining to fundamental programming and avoid serving as a medium for cheating by guiding students to solve their unique challenges. We will demonstrate our VTA system's initial implementation, which will be developed using Natural Language Processing (NLP) and deep learning approaches [3].

# LITERATURE REVIEW

Since ELIZA [4], an early chatbot, there have been significant advancements in chatbot technology. Using NLP and machine learning components, chatbots are artificially intelligent systems designed to simulate interactions with a single user or a group of users. Chatbots have both academic and commercial applications. In addition to non-academic applications, experts at Stanford University say that chatbots are better for students than other ways to talk [5]. The integration of education and AI will be aided by the introduction of big data, the constant emergence of digital campuses and online learning platforms [6]. Memon proposed an effective, simple, easy, and low-cost framework approach for designing a text-based multi-interactive chatbot. This will support the development of a multi-interactive chatbot's system for an educational area using AIML 2.0. This will also facilitate the students a personalized learning environment for their learning towards an outcome-based education domain [7]. Attempting to recall code line by line is not the proper method for comprehending a programming language [8], because students' learning capacities vary. Heller created Freudbot in 2005 to collect insights from psychology students in order to establish an environment in which students may participate and, therefore, overcome the difficulty of responding to questions [9]. Sadhasivam presented in his study "Implementation of Chatbot That Teach Programming Language " a chatbot named Progbot that can educate an underused programming language easily and intelligently [8]. Apprentices who are uncertain where to start can still inquire about the chatbot to start the lesson, and it begins with the fundamental topics and steadily advances to additional troublesome material. The request is replied to by the system's built-in artificial intelligence. The user has to choose the invalid reply button, which is able to alarm the system administrator [10], if they find any response that they are not looking for. The foremost related work to our paper could be A. Goel's presentation of Jill Watson [11]. Though it is not a personalized system [12], Jill Watson released the lecturers from responding to students' inquiries. The objective of this work is to provide a customized and intelligent virtual teaching assistant (VTA) chatbot designed to make students' learning experiences more fun and reduce the risk of failure. The VTA in ProTracer 2.0 will figure out if a student's answer is right or wrong by comparing it to the teacher's answer [13]. The extraction of intent and entities is a crucial stage in the process of conversation management. Rasa NLU is an adaptive mechanism that recognizes intent and entities in human speech [14]. The Rasa NLU model is passed to the NLU Interpreter after training, which decodes trial intents to determine whether NLU can correctly classify and retrieve entities [14]. A smart phone-based chatbot system named Alpha is executed in the Python programming language and uses the Dialog Flow framework, with a machine learning model to pull out the intents and a cloud-based database to store data [5]. The university's academic cloud stores lessons for students' mobile terminals. The VTA facilitates additional effective learner understanding by providing feedback instantly [15]. 'Coding Tutor' [16], 'e-Java' [17], 'ProgBot' [8] and 'PythonBot' [1] are available with affiliated literature. Regardless, most of them include extremely undersized content and concentrate on a particular programming language. This paper is related to existing literature to some extent still mainly focuses on support in the python programming language related courses for first-year Undergraduate students.

# PROBLEM STATEMENT

Our system detects the intents of the queries as a supervised classification problem where the set of intents, I = {"greetings", "method", "class", "loop", "if-else", ......"goodbye"}. Given a set of question patterns, T= {t1, t2, t3, t4, ...... tn} and their responses, R= {r1, r2, r3, r4, ...... rn} and other metadata information, our goal is to predict the correct intent label, Y= {y1, y2, y3, ...... yn} for generating the appropriate response. After processing the input set, our system generates a probability set, P= {p1, p2, p3, p4, ...... pn} through a matching algorithm, which represents the matching of queries with intents. The output with the maximum probability value indicates the correct intent [18].

# DATASET FOR VTA-BOT

## Primary Dataset Implementation

As there was no dataset available for fundamental python programming language course, we had to create a completely new dataset for this course. We took the help from some question-answering sites. Stack Overflow, Stack Exchange, Python.org, Programmers Heaven and Find-Nerd are among the question-answer sites for professional and hobbyist programmers. Again, the concise note supplied by our institution's lecturers and some available Python datasets gave the advantage to create a new dataset for the VTA-bot. At first, our objective was to create a flowchart of our work plan Fig.1. We started sample collection accordingly. At first, we chose ten books for the course which we were inspired of. Then, we formed a group of two members to choose books from among them. After selection, the whole group evaluated the books and agreed to gather samples. Next, we assigned the topic selection task to each member of the group. We initially compiled a list of chapter-specific topics. Then, we eliminated all repeated topics and compiled a list of unique ones. These topics encompass the entire course. One of the hardest tasks was to select questions and answers. After that, we distributed the unique topics to each member of the group. Then, each group member compiled all potential questions and answers for each topic. Lastly, the process of tagging the questions was one of the most challenging aspects of this study because arranging the questions according to a similar pattern, appeared to be perplexing. We gave a tag to every similar patterned question. Next, to verify the completeness of the data we obtained, we consulted a teacher's assistant for this course. According to the feedback we received, we modified our dataset. We included several essential questions acquired from the students' queries to our dataset. After fixing all the errors in the dataset, we finally obtained our desired dataset.

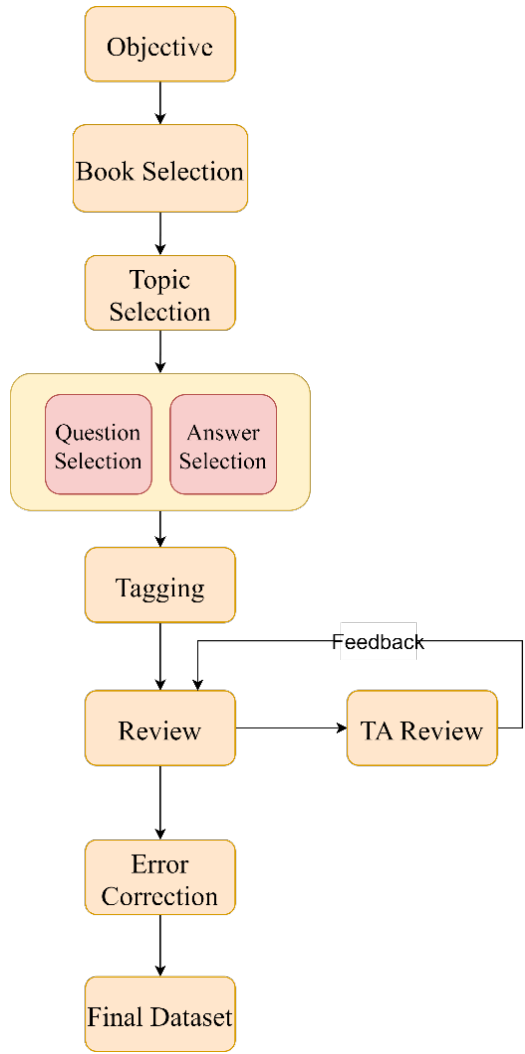

**FIGURE 1:** Work plan

## Data Hierarchy and Distribution

Our Python programming language data collection is organized hierarchically. The root is set to "intent". We saved "topics" inside of "intent". Each "topic" includes "tags". Again, each "tag" comprises questions of a similar kind. Lastly, each question corresponds to a certain response Fig.2. Currently, our dataset has a total of 214 unique tags, 1,006 questions, and 254 unique responses. Below are the percentages of questions associated with each category Fig.3. 18.7% of all tags have two or less questions, 18.2% contain three questions, 22.4% contain four questions, and the remainder contains five to ten questions. On average, each tag comprises between five to six questions.

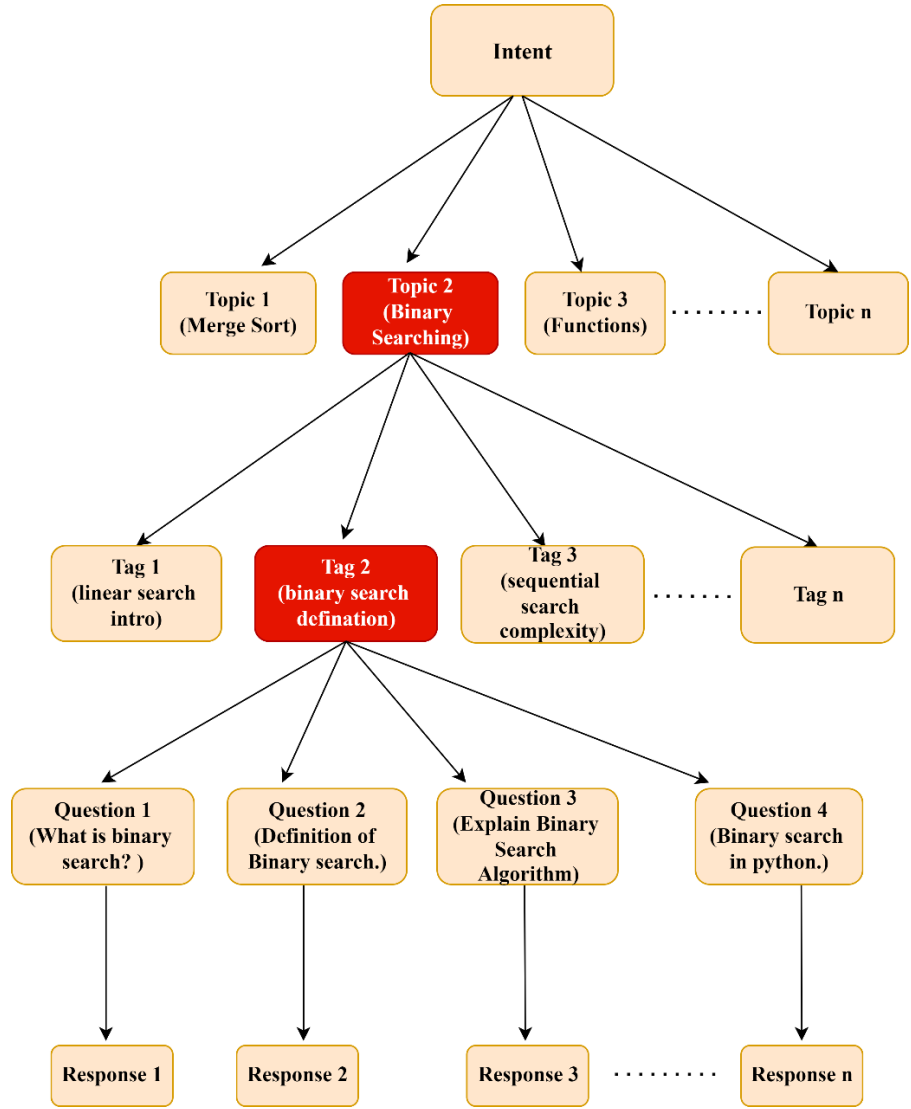

**FIGURE 2:** Data Hierarchy

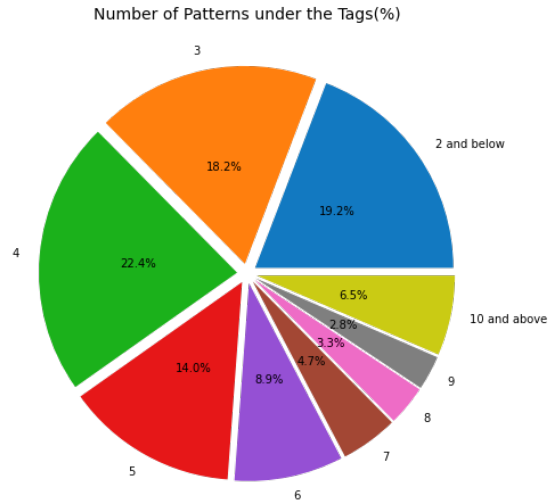

**FIGURE 3:** Number of Patterns Under the Tags (%)

## Data Pre-processing

To begin the preprocessing of our dataset, we must first import the dataset. The dataset was imported by copying it directly from the Google Sheet. We began with 1026 rows prior to removing the null rows. After removing rows containing null values from our dataset, we are left with 1006 rows. Then we performed some data preprocessing techniques, which include Case Folding, Punctuation Removal, Removal of Stopwords, Stemming, Lemmatization and Emoji Removal. At first, we converted all the values in our dataset to lowercase. After the lower-case conversion is complete, we removed the punctuation marks from the texts. After the punctuation has been eliminated, we then eliminated the stop words. Model training will be more accurate if the redundant words in the patterns are eliminated. For this we used a natural language processing approach called stemming. After all these approaches we made our data ready to fit on basic classifiers.

## Fitting dataset on basic classifiers

After all of the pre-processing, now we must evaluate how much efficient and balanced our dataset is. To determine this, we utilized four distinct classifiers and compared their respective accuracy gains. After evaluating the accuracy, we will be able to predict the accuracy of the neural net model of our prototype. Our chosen classifiers for training and testing are the Naive Bayes Classifier, the Decision Tree Classifier, the Linear Support Vector Machine, and the Logistic Regression classifier. After training and evaluating the models with the entire dataset, the accuracy for Naive Bayes Classifier is 21.42 percent, Decision Tree Classifier is 58.67 percent, Linear Support Vector Machine is 50 percent, and Logistic Regression is 62.24 percent Fig.4. Since, we were not entirely satisfied with the overall accuracy of the entire dataset, we omitted certain tags with less training patterns and refactored data while preserving the patterns of tags with at least 10 patterns. On rerunning these classifiers on the refactored data, the accuracy of each classifier improves. Fig.4 depicts the accuracy of each classifier for the refactored dataset. Now, the accuracy of Naive Bayes Classifier is 66.67 percent, Decision Tree Classifier is 75.75 percent, Linear Support Vector Machine is 69.70 percent and Logistic Regression is 75.75 percent. To get a clearer idea, we rerun all four classifiers using a more refactored dataset to determine if eliminating tags with fewer patterns yields a substantially higher degree of precision. Now, the accuracy for Naive Bayes Classifier is 73.91 percent, Decision Tree Classifier is 78.26 percent, Linear Support Vector Machine is 86.95 percent, and Logistic Regression is 82.60 percent Fig.4. With the more refactored dataset, we achieve the highest accuracy with Linear Support Vector Machine and Decision Tree Classifier Fig.4. Comparing the accuracy of all classifiers on the entire dataset, the refactored dataset, and the more refactored dataset in Fig.4, we can see that the more tags with a higher number of patterns we add, the more balanced our dataset becomes. After testing the accuracy, we can anticipate that the neural model of our prototype will have an accuracy of more than 90 percent. Our goal is to increase the accuracy of the entire dataset in the future by incorporating additional patterns for each tag, with the assistance of relevant faculties, student tutors, and also ourselves. In important evaluation metrics. It elegantly summarizes the predictive performance of a model by combining precision and recall. Initially, we test our four classifiers on the entire dataset and obtain an average f1-score Fig.5 of 0.16 for the Naive Bayes classifier, 0.56 for the Decision Tree classifier, 0.46 for the Linear Support Vector Machine (SVM), and 0.60 for the Logistic regression. Secondly, we retest all four classifiers on the refactored dataset and obtain an average f1-score Fig.5 of 0.64 for the Naive Bayes classifier, 0.75 for the Decision Tree classifier, 0.65 for the Linear Support Vector Machine (SVM), and 0.75 for the Logistic regression. When we finally retest all four classifiers on our more refactored dataset, we obtain an average f1-score Fig.5 of 0.71 for the Naïve Bayes classifier, 0.82 for the Decision Tree classifier, 0.85 for the Linear Support Vector Machine (SVM), and 0.81 for the Logistic regression classifier.

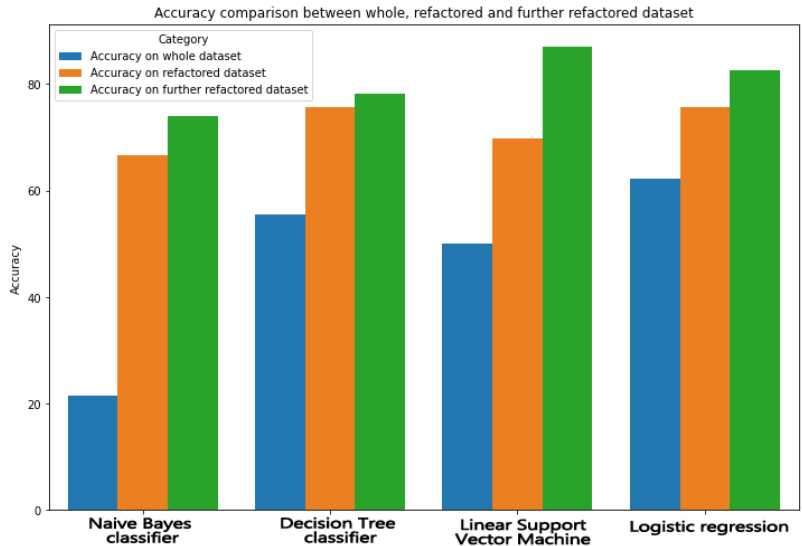

**FIGURE 4:** Accuracy comparison between whole, refactored and further refactored dataset

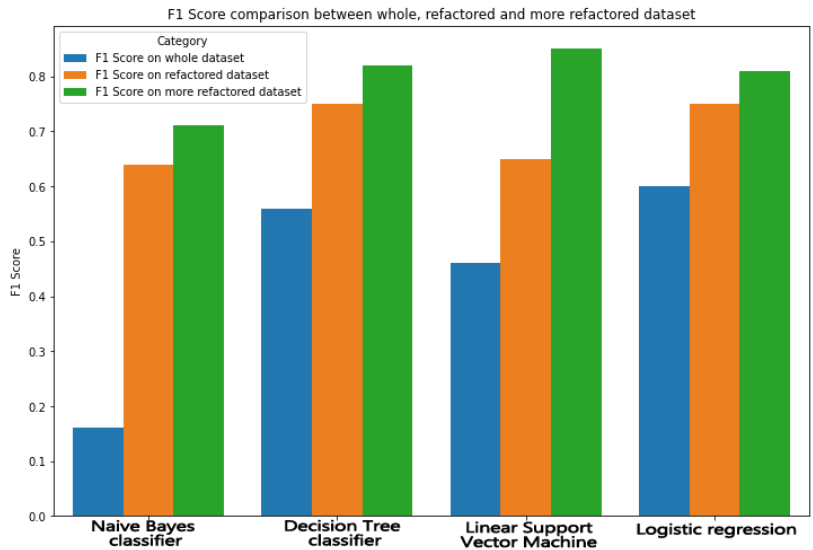

**FIGURE 5:** F1 score between whole, refactored and further refactored dataset

# PROTOTYPE DEVELOPMENT

## Methodology

We built our VTA-bot using python programming language following deep learning techniques. To begin, we installed PyTorch and NLTK. PyTorch is an open-source machine learning framework that accelerates the path from research prototyping to production deployment [19]. The Natural Language Toolkit (NLTK) is a platform used for building Python programs that work with human language data for application in statistical natural language processing (NLP) [34]. Following that, we created our training data set in a JSON file based on the course materials. We then implemented the NLP Utils. The NLTK module was used for this. We implemented the Neural Network model. PyTorch provided the torch.nn module to help us in creating and training the neural network model [19]. Then we created the Training Pipeline, which consists of multiple sequential steps that do everything from data extraction and preprocessing to model training and deployment. Finally, we implemented the virtual conversational agent by loading the trained model and making predictions for new sentences. We can modify the JSON dataset file with possible patterns and responses according to the course materials and re-run the training. In this project we built a custom AI-bot in Python and then integrated this agent to a website using Flask and JavaScript. Flask is a small and lightweight Python web framework that provides useful tools and features that make creating web applications in Python easier. In the Fig.6, we can see our VTA-bot is assisting a student. This VTA-bot is only a small portion of the VTA system we are proposing, and we are already developing other services for the system.

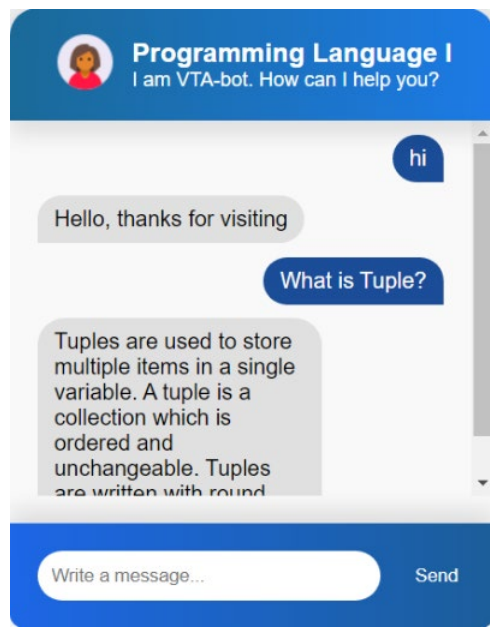

**FIGURE 6:** VTA-bot assisting a student

## NLP Preprocessing Pipeline

Data preprocessing is a machine learning data mining technique that converts raw data into a readable and understandable format. We prepared a data set which contains basic students' queries and their answers. We imported the dataset first, and then identified and handled missing values. We added intent for every similar patterned question and set the responses accordingly Fig.7.

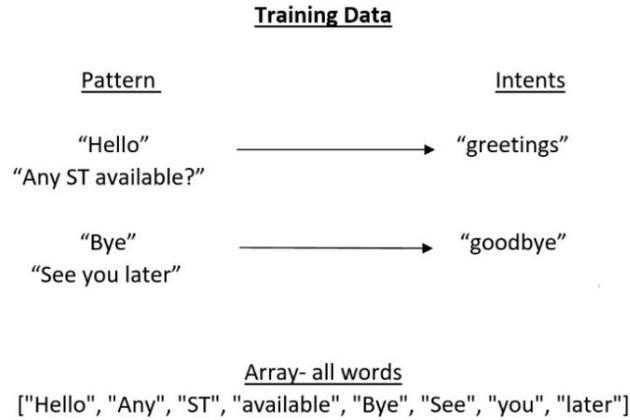

**FIGURE 7:** Data categorizing

We cannot just feed the input phrase to the neural network in its existing form. We need to translate the pattern strings into numbers that the network can interpret in some way. This is accomplished by transforming each statement into a "bag of words". To do this, we collected training words, or all the terms in the training data that our VTA-bot can look at and generated an array named "all words" Fig.7. The bag of words for each new sentence was then calculated using all of these keywords. A bag of words has the same size as the "all words" array, and each slot contains a 1 if the word appears in the incoming phrase, or a 0 if it does not. Prior to calculating the bag of words, we used two more natural language processing techniques: Tokenization and Stemming Fig.8.

## NLP Preprocessing Pipeline

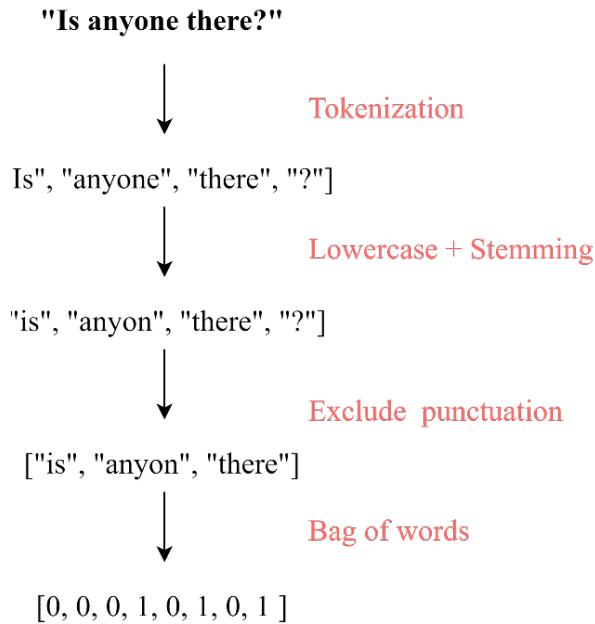

**FIGURE 8:** NLP Preprocessing Pipeline

## The Neural Network Model

A MLF neural network consists of neurons that are ordered into layers Fig.8. The first layer is called the input layer, the last layer is called the output layer, and the layers between are hidden layers. A Feed Forward Neural Net Fig.8 with two hidden layers is used to create our neural network model. Training and prediction are the two modes of operation supported by the Feed Forward neural network. Two data sets are required for the training and prediction of this neural network: the training set and the set to be predicted (test set). It takes the input, passes it through various layers one by one, and then finally generates a probability using a SoftMax function. This feed-forward Neural Network is constructed using the "torch.nn" package. The "torch.nn" is a python package provided by "PyTorch" that assisted us in developing and training the neural network.

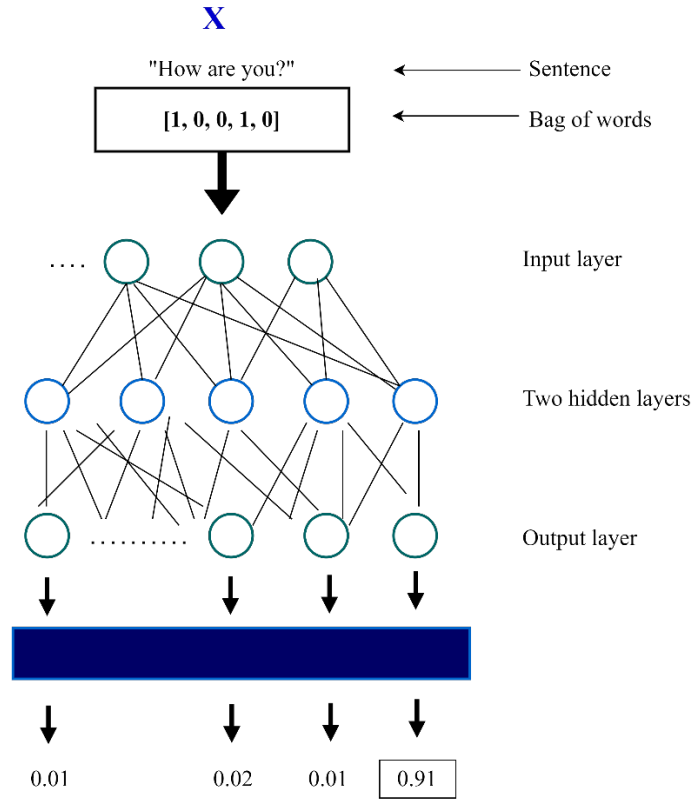

**FIGURE 9:** Feed Forward Neural Network Model

## RESULT

We load the dataset, which is in JSON format and after performing proper NLP techniques we generate the bag words for every pattern in the dataset. We trained our model with the updated data. We set the batch size to 8 and the number of epochs to 1,000. Then we calculated the loss and accuracy in each 100 epochs. The number of epochs is a hyper-parameter that controls how many times the learning algorithm runs over the whole training dataset. Loss is the error over the training set. A loss function takes the (output, target) pair of inputs and computes a value that estimates how far away the output is from the target. Throughout the training loop, in every epoch the loss is decreasing and the accuracy is increasing. In the Fig.11 we can see that, in the first iteration the loss was 1.1089. After our training concluded, our loss was effectively reduced to 0.0007. Again, in the Fig.10, we can see the accuracy increased to 95%.

Training the network is repeated for a large number of training set examples until the network finds a stable state. The most typical strategy for avoiding over-fitting is stopping early. Early stopping is based on separating data into training and validation sets and computing the validation error regularly during training. Training is terminated when the number of validation errors begins to rise.

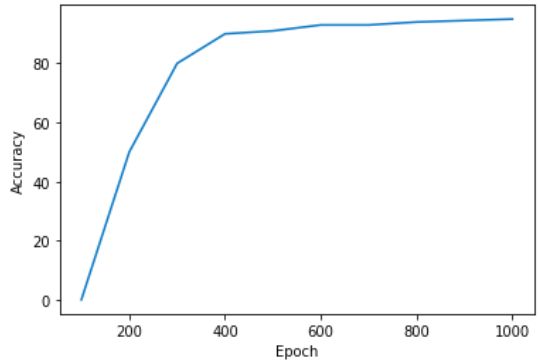

**FIGURE 10:** Accuracy Curve

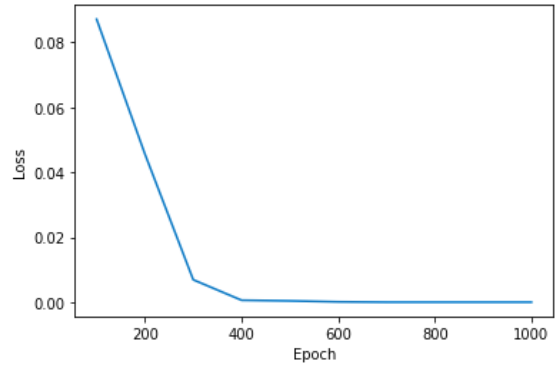

**FIGURE 11:** Loss Curve

## CONCLUSION

Students are sometimes hesitant to contact with strangers, and due to the newly established online-based education system, they get disoriented when transitioning to new programming concepts and lack adequate supervision. Our goal in writing this paper was to introduce our VTA-bot as a medium of learning to make their educational experiences more enjoyable and reduce their chances of failing. Along with answering core python programming questions, our VTA-bot will be able to assist students with course-related concerns and provide real-time assistance from their teachers. We demonstrated a preliminary implementation, which consists of a Chatbot that will assist and support students through the use of artificial intelligence (AI) technologies, including natural language processing (NLP). To help students even more, we intend to continue working on improving our system in the years to come, hoping for a revolution in our education sector.